# ANGULAR-MOMENTUM PROJECTION OF CRANKED SYMMETRY-UNRESTRICTED SLATER DETERMINANTS

H. ZDUŃCZUK
*Institute of Theoretical Physics, University of Warsaw,*
*ul. Hoża 69 Warsaw, Poland*
*zdunczuk@fuw.edu.pl*

J. DOBACZEWSKI
*Institute of Theoretical Physics, University of Warsaw,*
*ul. Hoża 69 Warsaw, Poland*
*dobaczew@fuw.edu.pl*

W. SATUŁA
*Institute of Theoretical Physics, University of Warsaw,*
*Hoża 69 Warsaw, Poland*
*satula@fuw.edu.pl*

We report on development of a new feature of the code HFODD, allowing for the angular-momentum projection of cranked symmetry-unrestricted Slater determinants. After a brief overview of the main theoretical building blocks and formalism, we present several preliminary applications. In particular, we discuss the case of a well-deformed rotational band in $^{156}$Gd, and we show the emergence of uncompensated poles in the overlap kernels calculated in an odd-$A$ nucleus $^{155}$Eu.

## 1. Introduction

Self-consistent mean-field method is practically the only formalism allowing for large-scale computations in heavy, open-shell nuclei with many valence particles. Inherent to the mean-field method is a mechanism of the spontaneous symmetry breaking, which is essentially the only way allowing to incorporate a significant part of many-body correlations into a single intrinsic (symmetry-violating) Slater determinant. Nuclear deformation and the emergence of rotational bands after applying simple cranked mean-field approach, i.e., after an approximate restoration of rotational invariance, is one of the most spectacular and most intuitive manifestations of the spontaneous symmetry breaking in nuclei.

An increasing quality of spectroscopic information demands more and more accurate theory, pushing the efforts beyond the mean-field approximation towards, in particular, symmetry-conserving formalisms and symmetry restoration.[1,2,3] Hereby, we report on development of a new theoretical tool allowing for the angular-momentum projection (AMP) of cranked symmetry-unrestricted Slater determinants.

The standard method used to develop symmetry-conserving theory, starting from *intrinsic* (deformed) wave function $|\Phi\rangle$, obtained within the mean-field approach, is provided by the *projection techniques* onto eigenspaces of symmetry operators. There are two practical realizations of the projection methods: more fundamental and elaborate *variation after projection* (VAP) and slightly less advanced *projection after variation* (PAV). In the past, many calculations have been performed within the angular-momentum PAV method, where non-rotating states have been projected, see, e.g., Refs.[4,5,6] and the reviews in Refs.[2,3]. The cranking method provides the first-order approximation to the angular-momentum VAP method.[1] However, after the ground-breaking studies in Refs.[7,8], calculations based on the AMP of cranked states have not been performed. Here, we present the first results of our recently developed AMP method of cranked symmetry-unrestricted Hartree-Fock states. The procedure we use has been implemented within the code HFODD (v2.25b).[9,10,11,12]

We determine the optimal product wave function $|\Phi\rangle$ by using the cranked self-consistent Skyrme-Hartree-Fock (SHF) method. The Ritz variational principle:

$$\delta \frac{\langle\Phi|\hat{H} - \omega \hat{I}_y|\Phi\rangle}{\langle\Phi|\Phi\rangle} = 0, \tag{1}$$

is equivalent to a variation of a local energy density functional (LEDF) with a supplementary constraint on the average value of the projection of angular momentum on the axis perpendicular to the symmetry axis ($y$ in our case). The variation of LEDF in the isoscalar-isovector $t = 0,1$ representation (potential part) can be formally written as:

$$\delta\mathcal{E}^{(Skyrme)} = \delta \sum_{t=0,1} \int d^3\mathbf{r}\, [\mathcal{H}_t^{(TE)}(\mathbf{r}) + \mathcal{H}_t^{(TO)}(\mathbf{r})]. \tag{2}$$

where $\mathcal{H}_t(\mathbf{r})$ is expressed as a bilinear form of the time-even (TE) $\rho, \tau, J$ and time-odd (TO) $\mathbf{s}$, $\mathbf{T}$, $\mathbf{j}$ local densities and currents, and by their derivatives,[13]

$$\mathcal{H}_t^{(TE)}(\mathbf{r}) = C_t^\rho \rho_t^2 + C_t^{\Delta\rho} \rho_t \Delta\rho_t + C_t^\tau \rho_t \tau_t + C_t^J J_t^2 + C_t^{\nabla J} \rho_t \nabla \cdot \mathbf{J}_t, \tag{3}$$

$$\mathcal{H}_t^{(TO)}(\mathbf{r}) = C_t^s \mathbf{s}_t^2 + C_t^{\Delta s} \mathbf{s}_t \Delta\mathbf{s}_t + C_t^T \mathbf{s}_t \cdot \mathbf{T}_t + C_t^j \mathbf{j}_t^2 + C_t^{\nabla j} \mathbf{s}_t \cdot (\nabla \times \mathbf{j}_t). \tag{4}$$

By taking an expectation value of the Skyrme force over the Slater determinant, one obtains the LEDF (3)–(4) with 20 coupling constants $C$ that are expressed uniquely through the 10 parameters $x_i, t_i, i = 0,1,2,3$, and $W, \alpha$ of the standard Skyrme force. Because of the local gauge invariance of the Skyrme force, only 14 coupling constants $C$ are independent quantities. The local gauge invariance links three pairs of time-even and time-odd constants in the following way:

$$C_t^j = -C_t^\tau, \qquad C_t^J = -C_t^T, \qquad C_t^{\nabla j} = C_t^{\nabla J}. \tag{5}$$

## 2. Angular-Momentum Projection

A deformed solution $|\Phi\rangle$ of Eq. (1) is a superposition of eigenstates of the angular-momentum operator. The angular-momentum conserving wave function can be obtained from the state $|\Phi\rangle$ by applying the angular-momentum projector:

$$|IMK\rangle = \hat{P}^I_{MK}|\Phi\rangle. \tag{6}$$

The AMP operator $\hat{P}^I_{MK}$ is a projector onto angular momentum $I$ with projection $M$ along the laboratory $z$ axis, and reads:[14,1]

$$\hat{P}^I_{MK} = \frac{2I+1}{8\pi^2}\int D^{I*}_{MK}(\Omega)\,\hat{R}(\Omega)\,d\Omega, \tag{7}$$

where $\Omega$ represents the set of three Euler angles $(\alpha,\beta,\gamma)$, $D^{I*}_{MK}(\Omega)$ are the Wigner functions,[15] and $\hat{R}(\Omega) = e^{-i\alpha\hat{I}_z}e^{-i\beta\hat{I}_y}e^{-i\gamma\hat{I}_z}$ is the rotation operator.

The state $|IMK\rangle$ is no longer a product wave function, but a complicated superposition of Slater determinants. The operator $\hat{P}^I_{MK}$ is not a projector in the mathematical sense.[1] It extracts from the intrinsic wave function the component with a projection $K$ along the intrinsic $z$ axis of the nucleus. Since $K$ is not a good quantum number, all these components must be mixed, and the mixing coefficients $g^{(i)}_K$ must be determined by the minimization of energy. This $K$-mixing is taken into account by assuming the following form for the eigenstates:

$$|IM\rangle^{(i)} = \sum_K g^{(i)}_K|IMK\rangle \equiv \sum_K g^{(i)}_K\hat{P}^I_{MK}|\Phi\rangle, \tag{8}$$

and by solving the following Hill-Wheeler (HW) equation:

$$\mathcal{H}\bar{g} = E\mathcal{N}\bar{g}, \tag{9}$$

where $\mathcal{H}_{K'K} = \langle\Phi|\hat{H}\hat{P}^I_{K'K}|\Phi\rangle$ and $\mathcal{N}_{K'K} = \langle\Phi|\hat{P}^I_{K'K}|\Phi\rangle$ denote the Hamiltonian and overlap kernels, respectively.

## 3. Application to $^{156}$Gd

We solve the cranked SHF equations for $^{156}$Gd by using the code HFODD for the SIII Skyrme-force parameters[16] and spherical harmonic-oscillator basis composed of $N_0$=10 shells. Then, we calculate the Hamiltonian and overlap kernels by using 50 Gauss-Chebyshev integration points in the $\alpha$ and $\gamma$ directions and 50 Gauss-Legendre points in the $\beta$ direction.[12] Figure 1 shows probability distributions $W_I$ of even angular-momentum components $I$ in the intrinsic cranked-SHF states $|\Phi_{I_y}\rangle$ constrained to $\langle\hat{I}_y\rangle = I_y$. The probability of finding the $I$ component can be calculated from:[1]

$$W_I = \sum_K\langle\Phi_{I_y}|\hat{P}^I_{KK}|\Phi_{I_y}\rangle. \tag{10}$$

The curves correspond to cranking wave functions with averaged angular momenta $\langle I_y\rangle = 0, 8, 12, 16,$ and $20\hbar$. One can see that for low angular momenta (e.g., for $I_y = 0$) the maxima of the distributions do not lie near the same values of $I \simeq \langle I_y \rangle$. Similar results have been obtained by Islam *et al.*[7] and Baye *et al.*[8].

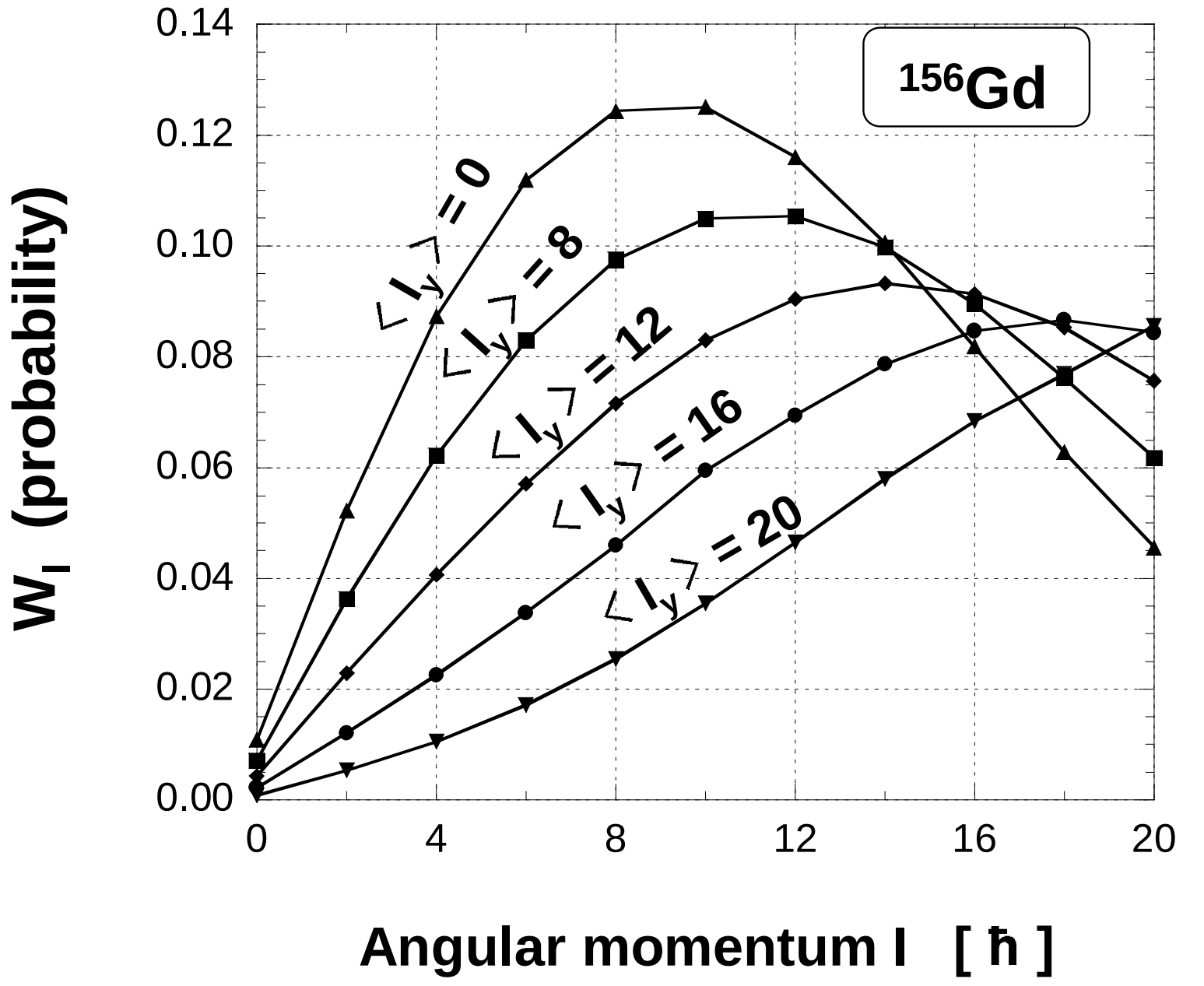

Fig. 1. Probability distributions of even angular-momentum components $I$ in $^{156}$Gd.

In Fig. 2 we show similar probability distributions $W_K$ of the projections $K$,

$$W_K = \langle \Phi_{I_y} | \hat{P}^I_{KK} | \Phi_{I_y} \rangle / W_I, \tag{11}$$

projected from the state with $\langle I_y \rangle = 10\hbar$ in $^{156}$Gd. One can see that the $K = 0$ component dominates for all angular momenta, while the $K > 0$ components increase with the increasing angular momentum. The $K = 1$ component is the second in magnitude after $K = 0$, which illustrates the build-up of the Coriolis coupling in a rotating intrinsic state. Note that only even values of $K$ can be projected from the non-rotating $\langle I_y \rangle = 0$ state, while all $K \geq 0$ appear in the cranked state.

In order to find energies of the AMP states, we solve Eq. (9) by diagonalizing first the norm matrix:

$$\mathcal{N}\bar{\eta} = n\bar{\eta}. \tag{12}$$

The non-zero eigenvalues ($n_m \neq 0$) of $\mathcal{N}$ are used afterwards to built the so-called *natural states*

$$|m\rangle = \frac{1}{\sqrt{n_m}} \sum_K \eta_m^{(K)} |IMK\rangle, \tag{13}$$

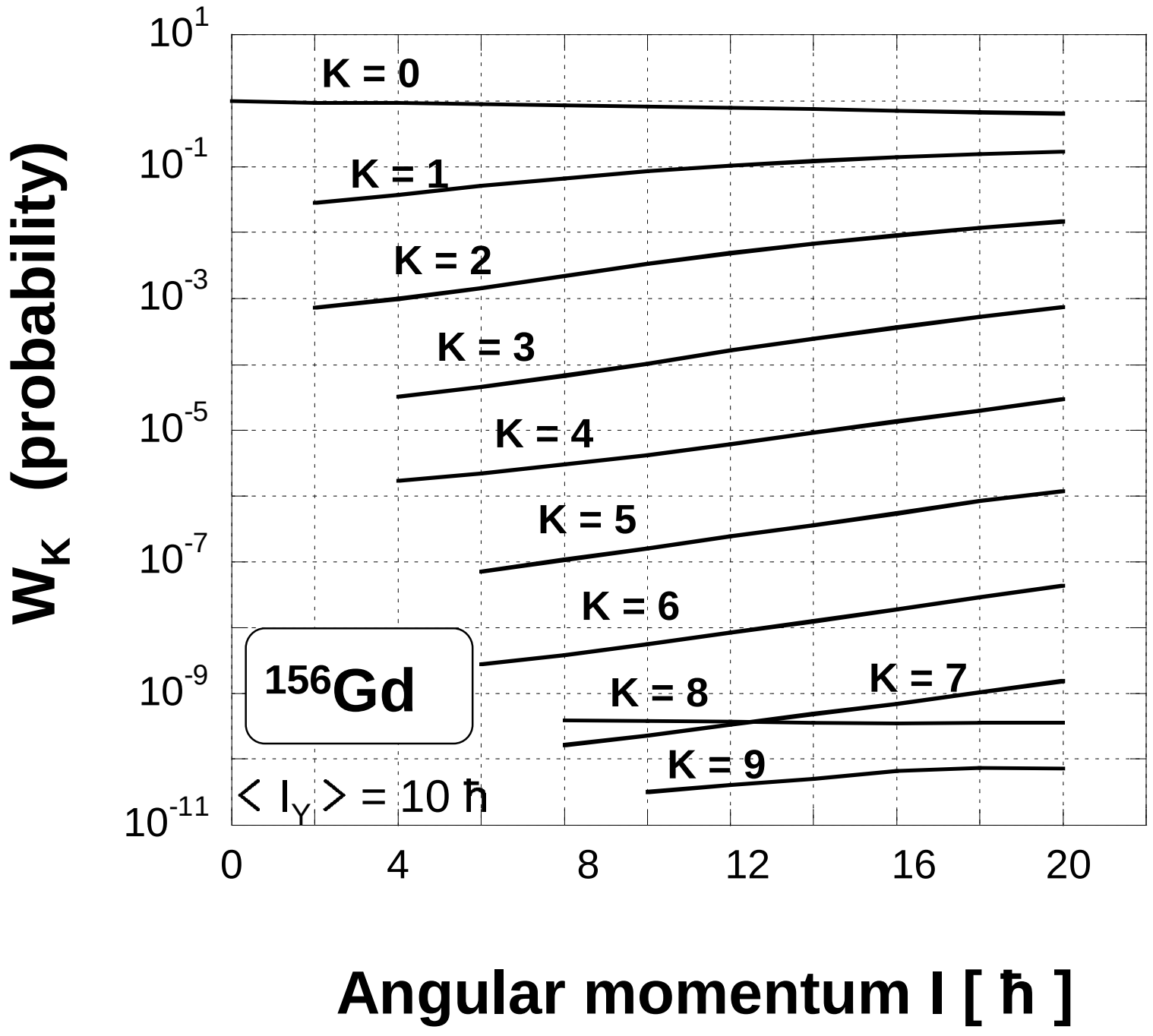

Fig. 2. Probability distributions of projections $K$ in the even angular-momentum components $I$ projected from the state with $\langle I_y \rangle = 10\hbar$ in $^{156}$Gd.

that span the subspace called *collective subspace*. Final diagonalization of the HW equation (9) is performed within the *collective subspace*. In this subspace the problem reduces to the standard hermitian eigenvalue problem.

In practical numerical applications we use the cut-off parameter $\zeta$ to construct the *collective subspace*, by keeping only the norm eigenstates (12) with $n_m \geq \zeta$. The test depicted in Fig. 3 shows the stability of projected solutions with respect to the number of the norm eigenstates kept in the collective subspace. The analysis was conducted for angular-momentum $0 \leq I \leq 20$ states projected from the deformed solution $|\Phi_{I_y=10}\rangle$ obtained by solving the cranked SHF equations with the constraint $\langle \Phi_{I_y=10}|\hat{I}_y|\Phi_{I_y=10}\rangle = 10\hbar$. The test clearly shows that, starting from a certain point, the obtained solutions are stable (plateau condition). Only for very small values of $\zeta < 10^{-8}$ the method becomes numerically unstable.

Figure 4 shows rotational bands (b)–(d) calculated in $^{156}$Gd in comparison with the experimental data (a). Band (b) represents the average mean-field energies obtained from the cranked SHF calculations by constraining solutions to $\langle I_y \rangle = I$. In bands (c) and (d), we show energies obtained by the AMP from the $\langle I_y \rangle = 0$ and $\langle I_y \rangle = I$, respectively. We see, that band (c) is much higher than bands (b) and (d), which indicates that the PAV from the $\langle I_y \rangle = 0$ state is not an adequate method of describing nuclear rotation. Note that the mean-field energies (b) and the AMP energies (d) corresponding to $\langle I_y \rangle = I$ turn out to be very similar to one another. This shows that the AMP from the $\langle I_y \rangle = I$ states constitutes a correct symmetry restoration of the approximate VAP solutions realized by the cranking procedure. The remaining discrepancy with experimental data must be attributed to pairing correlations, which are not included in our SHF solutions.

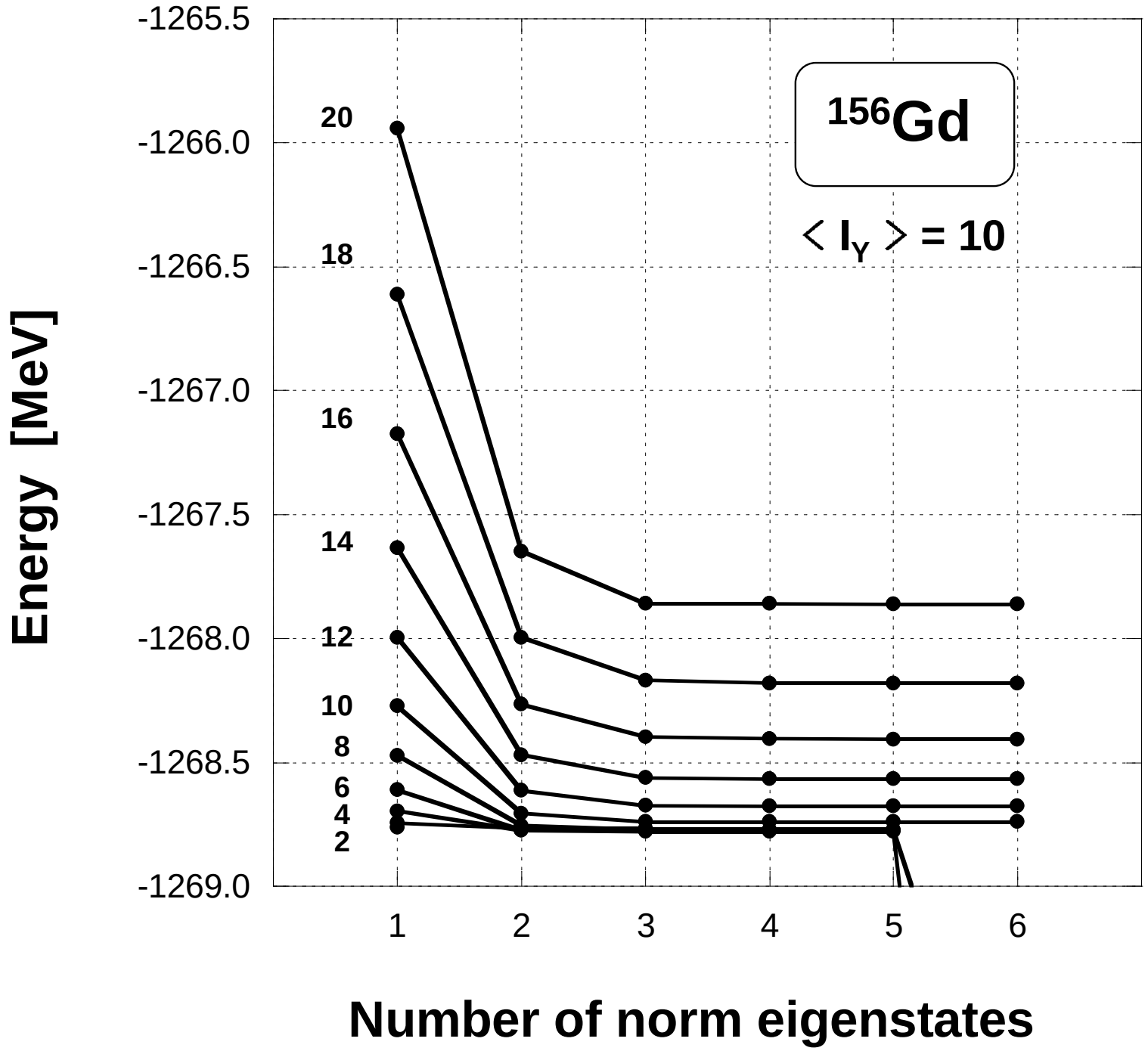

Fig. 3. Dependence of energies of projected states on the number of the norm eigenstates kept in the collective subspace. Angular momenta $0 \leq I \leq 20$ were projected from the state having the average projection of angular momentum $\langle \hat{I}_y \rangle = 10\hbar$.

When calculating the Hamiltonian kernels within the LEDF approach, one has to use transition density matrices between rotated states,

$$\rho_{\alpha\beta}(\Omega) = \frac{\langle \Phi | a^+_\beta a_\alpha \hat{R}(\Omega) | \Phi \rangle}{\langle \Phi | \hat{R}(\Omega) | \Phi \rangle}, \tag{14}$$

which are singular whenever the rotated states are orthogonal. In particular, the transition multipole moments,

$$Q_{\lambda\mu}(\Omega) = \sum_{\alpha\beta} (Q_{\lambda\mu})_{\beta\alpha} \rho_{\alpha\beta}(\Omega), \tag{15}$$

become singular for certain values of the Euler angles $\Omega$. This is illustrated in Fig. 5, which shows absolute values of the neutron and proton overlap kernels in

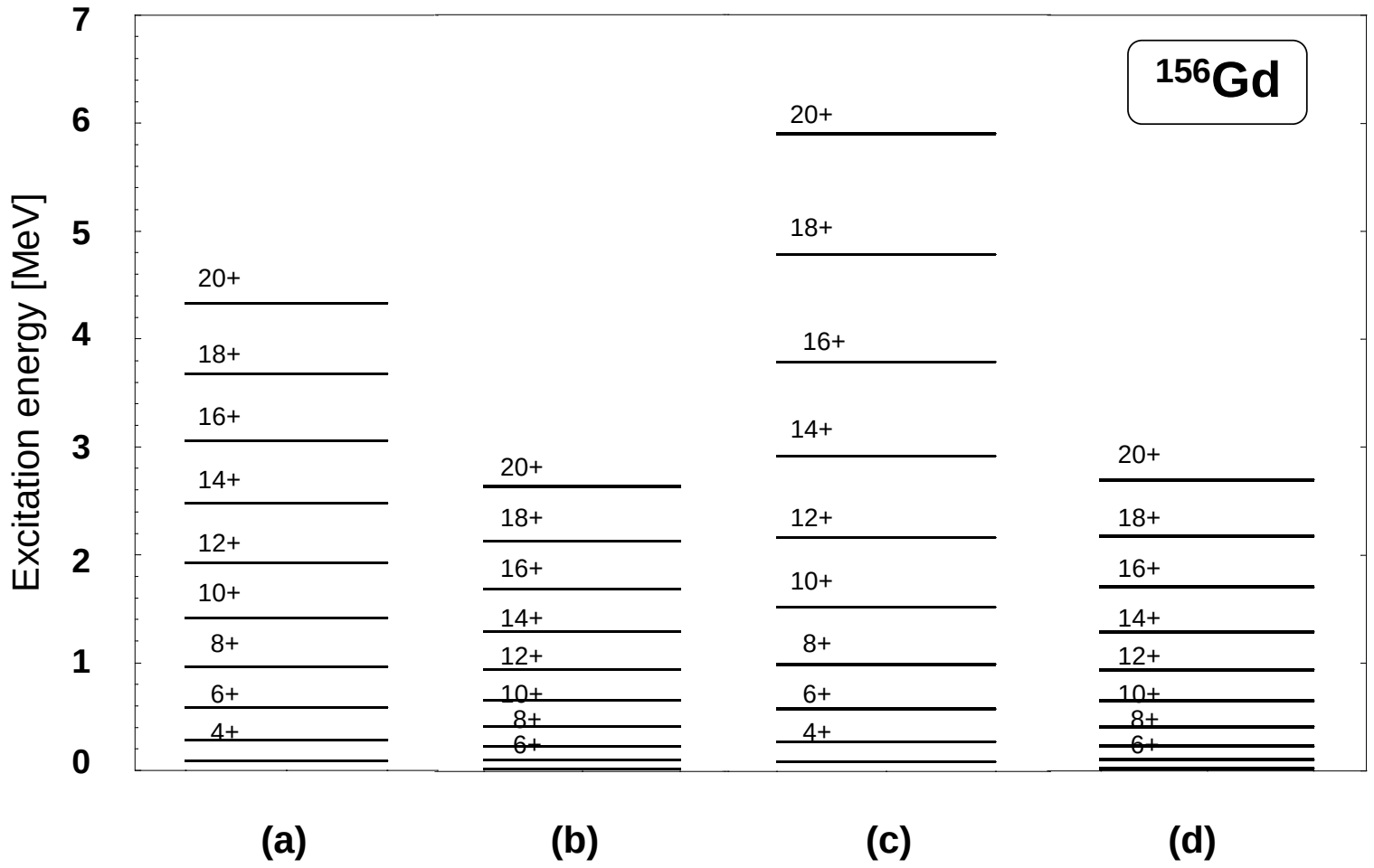

Fig. 4. Rotational bands in $^{156}$Gd nucleus: the four bands represent: (a) experimental data, (b) cranked SHF calculations, (c) band projected from the state $|\Phi_{I_y=0}\rangle$, and (d) band projected from the state $|\Phi_{I_y=I}\rangle$ into $I$, for $I = 0, 2, \ldots 20$.

$^{155}$Eu (left panel), and the transition quadrupole moments $Q_{20}(\Omega)$ in $^{155}$Eu and $^{156}$Gd (right panel). Calculations have been performed for axial shapes of nuclei, for which only the rotation about the Euler angle $\beta$ matters. The neutron overlap kernel corresponds to an even number of particles ($N = 92$), and is always positive, although at $\beta = 90^\circ$ it becomes as small as $10^{-14}$. On the other hand, the proton overlap kernel corresponds to an odd number of particles ($Z = 63$), and it three times changes the sign in the interval of $0^\circ \leq \beta \leq 180^\circ$. Consequently, the transition quadrupole moment of $^{156}$Gd is a regular function, while that of $^{155}$Eu has three poles.

Of course, when calculating the matrix elements of multipole operators between the rotated states, the transition matrix elements (15) are multiplied by the overlap kernels and the poles disappear. However, such a compensation is absent for kernels corresponding to higher powers of densities, viz. for the density-dependent terms of the Skyrme interactions, or for the direct-Coulomb-energy terms, or for the exchange-Coulomb-energy terms in the Slater approximation. How to treat such singular kernels within the AMP methods is currently an open and unsolved problem, similarly as is the case for the particle-number-projection methods recently discussed in Ref.[17].

## 4. Conclusions

We have developed a new theoretical tool allowing for the angular-momentum projection of cranked symmetry-unrestricted Slater determinants. In the present work, we presented preliminary applications pertaining to rotational bands in a well-deformed nucleus $^{156}$Gd. We showed that energies projected from non-rotating and rotating states are significantly different and similar, respectively, when compared to the cranking mean-field energies. We also showed that, due to vanishing overlaps between rotated states, transition energy kernels may have uncompensated poles in function of the Euler angles.

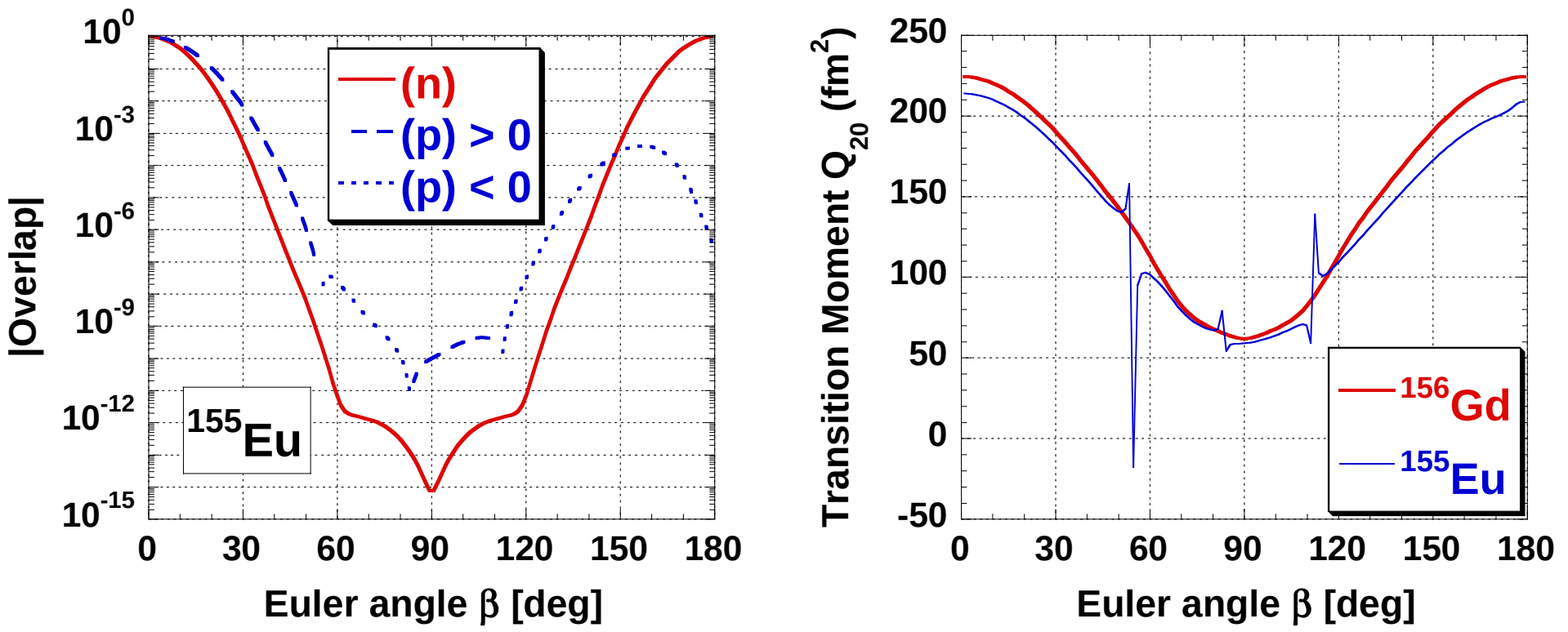

Fig. 5. Left panel: Dependence of the absolute value of the overlap between rotated states on the Euler angle $\beta$ in $^{155}$Eu. Solid line shows the neutron overlap that is always positive. Dashed and dotted lines show these segments of the proton overlap where it is positive and negative, respectively. Right panel: Transition quadrupole moments $Q_{20}$ versus $\beta$ in $^{156}$Gd (thick line) and $^{155}$Eu (thin line).

## Acknowledgments

This work was supported in part by the Polish Committee for Scientific Research (KBN) under contract No. 1P03B 059 27; and by the Foundation for Polish Science (FNP).